\documentclass[12pt]{article}
\usepackage{amsmath,amsthm,amscd,amssymb}
\usepackage{latexsym}
\theoremstyle{plain}

\theoremstyle{definition}

\theoremstyle{remark}

\newtheorem{case[theorem]}{Case}

\begin{document}

\title{GAUGE-DEPENDENT COSMOLOGICAL ``CONSTANT''}

\author{Bahram Mashhoon\\Department of Physics and
Astronomy\\University of Missouri\\Columbia,
Missouri 65211, U.S.A.\\Email: MashhoonB@missouri.edu \and Paul S.
Wesson\\Department of Physics\\University of Waterloo\\Waterloo,
Ontario N2L 3G1,
Canada\\Email: wesson@astro.uwaterloo.ca}
\maketitle

\begin{description}\item[PACs:] 0420, 0450, 1190, 9530
\item[Keywords:] Cosmological-Constant Problem, Higher-Dimensional Gravity,
Vacuum Fields, Cosmology and Galaxy Formation
\end{description}

\begin{abstract}
When the cosmological constant of spacetime is derived from the 5D
induced-matter theory of gravity, we show that a simple gauge transformation
changes it to a variable measure of the vacuum which is infinite at the big
bang and decays to an astrophysically-acceptable value at late epochs. \ We
outline implications of this for cosmology and galaxy formation.\end{abstract}

\section{Introduction}\baselineskip=20pt

In Einstein's theory of general relativity, the cosmological constant $
\Lambda $ is a fundamental parameter like the speed of light and the
gravitational constant. \ It measures the energy density of the vacuum, and
introduces a cosmological lengthscale of order $10^{28}$ cm based on
current astrophysical data \cite{1}. However, those same data imply that $
\Lambda $ could have been larger in the early universe, a possibility which
has been the subject of numerous investigations; see, for example,
\cite{New} for recent reviews of some of the phenomenological as well as
field-theoretical models involving variable cosmological
``constants''. This possibility can be addressed as well using
higher-dimensional gravitational theories. Extra dimensions have been
employed by many authors in connection with issues involving the
cosmological ``constant'' (see, for example, \cite{NR} and the references
cited therein). Such 
theories can also in principle help resolve the cosmological-constant
problem, which is basically the mismatch between the small value of $\Lambda
$ derived from cosmological observations and the large value it should have
as a measure of the vacuum fields of particle physics \cite{3}. 

      In this paper, we follow an approach based on the existence of
an extra spacelike dimension, and we will use the canonical formalism
of the induced-matter theory of gravity \cite{2} to  
show that a simple gauge transformation involving the extra coordinate of 5D
gravity changes $\Lambda $ so that it is infinite at the big bang and decays
to an astrophysically-acceptable value at the present time. This suggests
that $\Lambda $ is a gauge-dependent measure of the energy density of the
vacuum, opening the way to a potential resolution of the
cosmological-constant problem 
and helping with other astrophysical problems such as that of galaxy
formation.

The two current versions of 5D gravity theory are membrane theory and induced-matter
theory. In the former, gravity propagates freely (into the ``bulk''),
while the interactions of particle physics are confined to a hypersurface
(the ``brane''). Moreover, cosmological ``constants'' may exist both in
the bulk as well as on the brane. In the latter, there is no
restriction on the dynamics 
except that provided by the geodesic equation, and matter is explained as a
manifestation of the fifth dimension. The Ricci-flat condition is
imposed on the 5D manifold; therefore, the only possible cosmological
``constant'' is the one induced in 4D. Both theories involve conservation
laws couched in 5D terms, which perforce means that the 4D laws are
modified, resulting in a fifth force. The latter has been evaluated for
the induced-matter approach \cite{5} and the membrane approach \cite{6}, with
compatible results. Also, it is now known that the field equations for
these approaches are essentially the same \cite{7}. However, if we wish to
investigate the possibility that $\Lambda $ is a measure of the energy
density of a vacuum fluid, the most convenient formalism is the
induced-matter one. We will therefore adopt this below, extending previous
work on $\Lambda $ which indicated a connection to particle mass \cite{8} that
has recently been the subject of renewed interest \cite{9}. 

         The induced-matter theory in its simplest form is the basic
5D Kaluza-Klein theory in which the fifth dimension is not
compactified and the field equations of general relativity in 4D
follow from the fact that the 5D manifold is Ricci-flat; the large
extra dimension is thus responsible for the appearance of sources in
4D general relativity. In effect, the 4D world of general relativity
is embedded in a five-dimensional Ricci-flat manifold; indeed, this is
locally ensured by the Campbell theorem \cite{Cam}. We assume in what
follows that the extra dimension is spacelike.

An interesting result of the induced-matter theory is that if $ds^2 =
g_{\mu \nu} (x) \,dx^\mu \,dx^\nu$ is the 4D metric of any matter-free
spacetime 
with a cosmological constant $\Lambda > 0$ in general relativity, then
$dS^2 = ( l / L )^2\, ds^2 - dl^2$ with $L^2 = 3 / \Lambda$ is the metric of
a 5D manifold that is Ricci-flat \cite{8}. Conversely, any 5D Ricci-flat
metric of this canonical form corresponds to a matter-free spacetime
with metric $ds^2$ and an effective cosmological constant $\Lambda = 3 /
L^2$. We are particularly interested in the question of the uniqueness
of the latter correspondence. An important example is provided by the
de Sitter solution of inflationary cosmology. In view of its basic
significance, we first concentrate on this solution that is
conformally flat in 4D and we write its metric tensor in the form $g_{\mu
\nu} (x) = k(x) \eta_{\mu\nu}$. The explicit form of $k(x)$ is of no
consequence for our discussion, since we are using the simple case of
de Sitter spacetime to illustrate a rather general result. The
question is then whether from the 5D standpoint the Ricci-flat metric
$dS^2 = ( l / L )^2 f( x, l ) \eta_{\mu\nu}\, dx^\mu \,dx^\nu - dl^2$
has a unique 
solution for the function $f$. Clearly $f = k(x)$ is a possible solution,
but it may not be the only solution. In fact, we find that in general
$f = ( 1 - l_0 / l )^2 k(x)$, where $l_0$ is an arbitrary constant. For
$l_0 = 0$, we recover $f(x, l) = k(x)$; however, a novel situation arises
in the generic case that $l_0 \neq 0$. This paper is devoted to a
detailed derivation, interpretation and generalization of this result.

We work in 5D for the sake of simplicity; moreover, 5D theories are
widely regarded as the low-energy limit of even 
higher-dimensional theories \cite{10}. These include 10D supersymmetry, 11D
supergravity and 26D string theory. These theories hold out the hope of
unifying gravity with the interactions of particle physics, but our aim in
what follows is to lay a solid 5D foundation.

The prospect of going from $\Lambda $ = constant to $\Lambda $ = $\Lambda $
(time, space) is an intriguing one \cite{Man}. However, it is also a fundamental
shift from the way this parameter is viewed in general relativity.
Therefore, in the next section we will not skimp the details of how
we go
from a constant $\Lambda $ to a time-variable one; and we will be careful
with our comments about extending this to the space-variable case. A
fundamental rethink of $\Lambda $ can be justified for any
higher-dimensional theory: the group of gauge changes (coordinate
transformations) of the bigger space will necessarily affect the physics of
the smaller space, if the change involves the higher coordinates. We
will show how this works for a simple case in Section 
2. Those more interested in physics than mathematics will find a
summary of our
results in Section 3.

\section{A 5D Gauge Transformation that Changes\newline
the 4D Cosmological ``Constant''}

We start with the 5D canonical metric of the induced-matter theory of
gravity. 
Indeed, we will draw on previous work \cite{2,8} and
use the same notation. (Lower-case Greek letters will run 0, 123 for time
and space. Upper-case English letters will run 0, 123, 4 with $x^{4}=l$ as
the extra coordinate. Geometrical units will render the speed of light and
the gravitational constant both unity.) Our aim is to show that the simple
gauge transformation $l\rightarrow \left( l-l_{0}\right) $ changes the
structure of the field equations significantly, taking the cosmological
constant $\Lambda $ from a true constant to an $l$-dependent
parameter. The analysis will prove to be nontrivial (despite the
simple nature of the
gauge change). We will later confirm the result for $\Lambda \rightarrow
\Lambda \left( l\right) $ by a less informative but quicker method.

The line element for the canonical metric can be written \cite{8} in the form
\begin{equation}
dS^{2}=\frac{l^{2}}{L^{2}}\left[ g_{\alpha \beta }\left( x^{\gamma
},l\right) dx^{\alpha }dx^{\beta }\right] -dl^{2}\;\;\;.
\end{equation}
This 5D element contains the 4D one $ds^{2}=g_{\alpha \beta }\left(
x^{\gamma },l\right) dx^{\alpha }dx^{\beta }$. The $l$-dependence of the
4D metric tensor is necessary in order to preserve generality, since (1)
uses all of the 5 available degrees of coordinate freedom to set the
electromagnetic potentials $\left( g_{4\alpha }\right) $ to zero and to set
the scalar potential $\left( g_{44}\right) $ to a constant. In general,
our 4D physics takes place on a hypersurface of (1) specified by a value of $
l$, about which particles do not wander freely but are constrained by the 5D
geodesic equation (see below and refs. 6, 7, 10). The signature of (1) is $
\left( +----\right) $, since we have assumed a spacelike extra dimension. For
this choice, the constant $L$ in (1) is related to the cosmological constant
$\Lambda $ via $\Lambda =3/L^{2}$; specifically, if $\partial g_{\alpha \beta}
/ \partial l = 0$, then the Ricci-flat requirement in 5D reduces to $R_{\alpha
\beta} = \Lambda g_{\alpha \beta}$ in 4D. This result has been known for a
decade, and follows from the field equations. The latter in terms of the
Ricci tensor are $R_{AB}=0\left( A,B=0,\;123,\;4\right) $. These 15
relations can always be written as 1 wave equation, 4 conservation
equations, and 10 Einstein equations \cite{2}. The latter in terms of the
Einstein tensor and the energy-momentum tensor are $G_{\alpha \beta }=8\pi
T_{\alpha \beta }\left( \alpha ,\beta =0,123\right) $. The source tensor
may contain parts due to ``ordinary'' matter and parts due to the
``vacuum'', and we will see below that the second of these depends
critically on the fifth dimension. That is, in the {\it general}
case, the physics of the 4D vacuum which follows from (1) depends critically
on the choice of the 5D gauge.

Here, we look at a {\it special} but physically-instructive case of
(1). That metric is general, so to make progress we need to apply some
physical filter to it. Now, the physics of the early universe is commonly
regarded as related to inflation; and the standard 4D metric for this is
that of de Sitter, where $ds^{2}=dt^{2}-\exp \left[ 2\sqrt{\Lambda \diagup
3\,}t\right] d\sigma ^{2}$. (Here $d\sigma ^{2}\equiv dr^{2}+r^{2}d\theta
^{2}+r^{2}\sin ^{2}\theta \,d\phi ^{2}$ in spherical polar
coordinates.) The physics flows essentially from the cosmological
constant $\Lambda $. However, it is
well known that the de Sitter metric is conformally flat. This
suggests that physically-relevant results in 4D may follow from the
metric (1) in 5D if the latter
is restricted to the conformally-flat form:

\begin{equation}
dS^{2}=\frac{l^{2}}{L^{2}}\left[ f\left( x^{\gamma },l\right) \eta _{\alpha
\beta }dx^{\alpha }dx^{\beta }\right] -dl^{2}\;\;\;\;\;.
\end{equation}
Here $\eta _{\alpha \beta }=$ diagonal $\left( +1-1-1-1\right) $ is the
metric for flat Minkowski space. We 
are particularly interested in the $l$-dependence of
$f\left(
x^{\gamma },l\right) $. To determine the latter, we need to solve the
field equations.

The components of the 5D Ricci tensor for the general metric (1) are
\begin{subequations}\begin{align}
^{\left( 5\right) }R_{55} &=-\frac{\partial A_{\,\,\alpha }^{\alpha }}{
\partial l}-\frac{2}{l}A_{\,\,\alpha }^{\alpha }-A_{\alpha \beta }A^{\alpha
\beta },  \\
^{\left( 5\right) }R_{\mu 5} &=A_{\mu}{}^{\,\alpha }{}_{;\alpha}-\frac{
\partial \Gamma _{\mu \alpha }^{\alpha }}{\partial l},   \\
^{\left( 5\right) }R_{\mu \nu }
&=^{\left(4\right)}\hspace*{-4pt}R_{\mu \nu }-S_{\mu \nu }\;\;\;\;,
\end{align}\end{subequations}
where $S_{\mu \nu }$ is a symmetric tensor given by
\begin{equation}
S_{\mu \nu }\equiv \frac{l^{2}}{L^{2}}\left[ \frac{\partial A_{\mu \nu }}{
\partial l}+\left( \frac{4}{l}+A_{\,\,\alpha }^{\alpha }\right) A_{\mu \nu
}-2A_{\mu }{}^{\alpha }A_{\nu \alpha }\right] +\frac{1}{L^{2}}\left(
3+lA_{\,\,\alpha }^{\alpha }\right) g_{\mu \nu \;\;\;\;.}
\end{equation}
Here $^{\left( 4\right) }R_{\mu \nu }$ and $\Gamma _{\nu \rho }^{\mu }$ are,
respectively, the 4D Ricci tensor and the connection coefficients
constructed from $g_{\alpha \beta }$. Moreover
\begin{equation}
A_{\alpha \beta }\equiv \frac{1}{2}\frac{\partial g_{\alpha \beta }}{
\partial l}\;\;\;\;,
\end{equation}
where $A_{\alpha }{}^{\beta }=g^{\beta \delta }A_{\alpha \delta }$, and the
semicolon in equation (3b) represents the usual 4D covariant
derivative. We need to solve (3) in the form $R_{AB}=0$, subject to
putting $g_{\mu \nu
}\left( x^{\gamma },l\right) =f\left( x^{\gamma },l\right)
\eta_{\mu\nu}$ as in (2),
which ensures (4D) conformal flatness. We note that $g^{\mu \nu }=\eta
^{\mu \nu }\diagup f$ and $A_{\mu \nu }=f^{\prime }\eta _{\mu \nu }\diagup 2$,
where $f^{\prime }\equiv \partial f\left( x^{\gamma },l\right) \diagup
\partial l$. Also, $A^{\alpha \beta }=f^{\prime }\eta ^{\alpha \beta
}\diagup (2f^{2}),\;A_{\;\;\alpha }^{\alpha }=2f^{\prime }\diagup f$ and $
A_{\;\;\beta }^{\alpha }=f^{\prime }\eta _{\;\;\beta }^{\alpha
}\diagup (2f).$ Then the scalar component of the field equation (3a)
becomes
\begin{equation}
2\frac{\partial}{\partial l}\left( \frac{f^{\prime }}{f}\right) +\left( \frac{
f^{\prime }}{f}\right) ^{2}+\frac{4}{l}\left( \frac{f^{\prime }}{f}\right)
=0\;\;\;\;.
\end{equation}
To solve this, we define $U\equiv f^{\prime }\diagup f+2\diagup l$. Then
(6) is equivalent to $2U^{\prime }+U^{2}=0$, or $\partial \left(
U^{-1}\right) \diagup \partial l=1\diagup 2$, so on introducing an arbitrary
function of integration $l_{0}=l_{0}\left( x^{\gamma }\right) $ we obtain $
U^{-1}=\left[ l-l_{0}\left( x^{\gamma }\right) \right] \diagup 2$. This in
terms of the original function $f$ means that $f^{\prime }\diagup f+2\diagup
l=U=2\diagup \left[ l-l_{0}\left( x^{\gamma }\right) \right] $, or $\partial
\left[ \ln \left( l^{2}f\right) \right] \diagup \partial l=\partial \left\{
\ln \left[ l-l_{0}\left( x^{\gamma }\right) \right] ^{2}\right\} \diagup
\partial l$. This gives $l^{2}f\diagup \left[ l-l_{0}\left( x^{\gamma
}\right) \right] ^{2}=k\left( x^{\gamma }\right) $, where $k=k\left(
x^{\gamma }\right) $ is another arbitrary function of integration. We have
noted this working to illustrate that the solution of the scalar component
of the field equations (3a) or (6) involves an arbitrary length $l_{0}\left(
x^{\gamma }\right) $ and an arbitrary dimensionless function $k\left(
x^{\gamma }\right) $. To here, the solution for the conformal factor in
the metric $g_{\mu \nu }\left( x^{\gamma },l\right) =f\left( x^{\gamma
},l\right) \eta _{\mu \nu }$ is
\begin{equation}
f\left( x^{\gamma },l\right) =\left[ 1-\frac{l_{0}\left( x^{\gamma }\right)
}{l}\right] ^{2}k\left( x^{\gamma }\right)
\end{equation}
and involves both arbitrary functions.

However, one of these is actually constrained by the vector component of the
field equations (3b). To see this we note that $A_{\mu \nu }$ of (5) is
symmetric, and it is a theorem that then
\begin{equation}
A_{\nu\;\; ;\mu }^{\;\;\mu }=A_{\;\;\nu ;\mu }^{\mu }=\frac{1}{\sqrt{-g}}
\frac{\partial }{\partial x^{\mu }}\left( \sqrt{-g}A_{\;\;\nu }^{\mu
}\right) -\frac{A^{\alpha \beta }}{2}\frac{\partial g_{\alpha \beta }}{
\partial x^{\nu }}\;\;\;.
\end{equation}
Here $g$ is the determinant of the 4D metric, so since $g_{\mu \nu }=f\eta
_{\mu \nu }$ we have $\sqrt{-g}=f^{2}$. Then using (8), equation (3b)
becomes
\begin{equation}
\frac{1}{2f^{2}}\frac{\partial }{\partial x^{\mu }}\left( f\,f^{\prime
}\,\delta _{\;\;\nu }^{\mu }\right) -\frac{f^{\prime }}{f^{2}}\frac{\partial
f}{\partial x^{\nu }}=\frac{\partial }{\partial l}\left( \Gamma _{\nu
\;\alpha }^{\alpha }\right) \;\;\;\;.
\end{equation}
The right-hand side of this can be re-expressed using the identity $\Gamma
_{\nu \;\alpha }^{\alpha }\equiv \left( \sqrt{-g}\right) ^{-1}\partial
\left( \sqrt{-g}\right)$ $\diagup \partial x^{\nu }$, whence (9) becomes
\begin{equation}
\frac{1}{2f^{2}}\frac{\partial }{\partial x^{\nu }}\left( f\,f^{\prime
}\right) -\frac{f^{\prime }}{f^{2}}\frac{\partial f}{\partial x^{\nu }}=2
\frac{\partial }{\partial l}\left[ \frac{f}{f^{2}}\frac{\partial f}{\partial
x^{\nu }}\right] \;\;\;\;.
\end{equation}
In this form, we can multiply by $2f^{2}$ and re-arrange to obtain
\begin{equation}
f\frac{\partial f'}{\partial x^{\nu }} =f'
\frac{\partial f}{\partial x^{\nu }}\;\;\;\;.
\end{equation}
Dividing by $f\,f'\neq 0$, we find
\begin{equation}
\frac{\partial }{\partial x^{\nu }}\left( \frac{f'}{f}\right)
=0\;\;\;\;.
\end{equation}
But the term in parenthesis here, by (7), is $f'\diagup
f=2l_{0}\left( x^{\gamma }\right) \diagup \{l\left[ l-l_{0}\left( x^{\gamma
}\right) \right]\} $. Thus (12) implies that $l_{0}\left( x^{\gamma }\right)
=l_{0}$ and is constant. We have noted this working to illustrate that the
scalar and vector components of the field equations (3a) and (3b) together
yield the conformal factor
\begin{equation}
f\left( x^{\gamma },l\right) =\left( 1-\frac{l_{0}}{l}\right) ^{2}k\left(
x^{\gamma }\right) \;\;\;\;,
\end{equation}
which involves only one ``arbitrary'' function that is easy to
identify: if the constant parameter $l_0$ vanishes, then $k \eta_{\mu
\nu}$ is
simply our original de Sitter metric tensor. 

The tensor component of the field equations (3c) does {\it not}
further constrain the function $k\left( x^{\gamma }\right) $. However, we need to
work through this component in order to isolate the 4D Ricci tensor $
^{\left( 4\right) }R_{\mu \nu }$ and so obtain the effective cosmological
constant. To do this, we need to evaluate $S_{\mu \nu }$ of (4). The
working for this is straightforward but tedious. The result is simple,
however:
\begin{equation}
S_{\mu \nu }=\frac{3}{L^{2}}k\left( x^{\gamma }\right) \eta _{\mu \nu
\;\;\;\;.}
\end{equation}
By the field equations (3c) in the form $^{\left( 5\right) }R_{\mu \nu }=0$,
this means that the 4D Ricci tensor is also equal to the right-hand side of
(14). We recall that our (4D) conformally-flat spaces (2) have $g_{\mu \nu
}=f\left( x^{\gamma },l\right) \eta _{\mu \nu }=\left( 1-l_{0}\diagup
l\right) ^{2}k\left( x^{\gamma }\right) \eta _{\mu \nu }$ using (13)
above. Thus $k\left( x^{\gamma }\right) \eta _{\mu \nu }=l^{2}g_{\mu
\nu }\diagup
\left( l-l_{0}\right) ^{2}$ and
\begin{equation}
^{\left( 4\right) }R_{\mu \nu }=\frac{3}{L^{2}}\frac{l^{2}}{\left(
l-l_{0}\right) ^{2}}g_{\mu \nu }\;\;\;\;.
\end{equation}
This is equivalent to the Einstein field equation for the de Sitter
metric tensor $k \eta_{\mu \nu}$, since under a constant conformal scaling of
a metric tensor, the corresponding Ricci tensor remains
invariant. None the less, (15) defines an Einstein space $^{\left( 4\right) }R_{\mu \nu }=\Lambda
g_{\mu \nu }$ with an effective cosmological constant given by
\begin{equation}
\Lambda =\frac{3}{L^{2}}\left( \frac{l}{l-l_{0}}\right) ^{2}\;\;\;\;.
\end{equation}
This is our main result; for $l_0 = 0$, $\Lambda$ reduces to the de
Sitter cosmological constant. 

It differs from the ``standard'' one $\Lambda =3\diagup L^{2}$, which is
obtained by reducing the 5D field equations to the 4D Einstein equations for
a pure-canonical metric in which the 4D metric tensor does not depend on the
extra coordinate $l$ \cite{8}. The difference between the results is
mathematically modest, but can be physically profound, because (16) admits
the possibility that $\Lambda \rightarrow \infty $ for $l\rightarrow
l_{0}$. We will return to this below, but here we wish to make some
comments about
the nature of (16).

The (4D) conformally-flat metric we are considering here and the
pure-canonical metric considered by other workers [5, 6, 8--10]
have 5D line 
elements given respectively by
\begin{subequations}\begin{align}
dS^{2} &=\frac{\left( l-l_{0}\right) ^{2}}{L^{2}}k\left( x^{\gamma }\right)
\,\eta _{\alpha \beta }\,dx^{\alpha }dx^{\beta }-dl^{2},   \\
dS^{2} &=\frac{l^{2}}{L^{2}}g_{\alpha \beta }\left( x^{\gamma }\right)
dx^{\alpha }dx^{\beta }-dl^{2}\;\;\;\;.
\end{align}\end{subequations}
Clearly the two are compatible, and the second implies the first if we shift
$l\rightarrow \left( l-l_{0}\right) $ and write $g_{\alpha \beta }\left(
x^{\gamma }\right) =k\left( x^{\gamma }\right) \eta _{\alpha \beta }$. Of
course, we can always make the (apparently trivial) coordinate
transformation or gauge change $l\rightarrow \left( l-l_{0}\right) $. This
leaves the extra part of the canonical metric (17b) unchanged, while the
prefactor on the 4D part changes from $l^{2}\diagup L^{2}$ to $\left(
l-l_{0}\right) ^{2}\diagup L^{2}=\left( l^{2}\diagup L^{2}\right) \left[
\left( l-l_{0}\right) \diagup l\right] ^{2}$. 

Let us now replace $k(x^\gamma) \eta_{\alpha\beta}$ in (17a) by a generic
metric tensor $g_{\alpha \beta} ( x^\gamma)$  and write $\overline{g}_{\alpha 
\beta }=\left[ \left( l-l_{0}\right) \diagup l\right] ^{2}g_{\alpha \beta }$
. Then we obtain a line element which looks like (17b) except that $
g_{\alpha \beta }$ has been replaced by $\overline{g}_{\alpha \beta
}$. Now it is a theorem that solutions of the source-free 5D field
equations $
R_{AB}=0$ with metric (17b) satisfy the source-free 4D field equations $
R_{\alpha \beta }=\Lambda g_{\alpha \beta }$ with $\Lambda =3\diagup L^{2}$
\cite{8}. Therefore, the same must hold with $R_{\alpha \beta }=
R_{\alpha \beta }\left( \overline{g}_{\alpha \beta }\right) =
\overline{\Lambda }\overline{g}_{\alpha \beta }$ and $\overline{\Lambda }
=\left( 3\diagup L^{2}\right) \,l^{2}\diagup \left( l-l_{0}\right)
^{2}$, since a constant conformal transformation of the metric leaves the Ricci
tensor invariant. This is identical to (16) above. Put another way: A
translation along the $ 
l$-axis preserves the form of the canonical metric, and since the 5D field
equations are covariant we obtain again the 4D field equations, but with a
different cosmological constant.

This is an elementary example of a situation that has been alluded to before
in the literature \cite[p. 125]{2}: 5D quantities $Q=Q\left( x^{A}\right) $
are preserved under $x^{A}\rightarrow \overline{x}^{A}\left( x^{B}\right) $,
but 4D quantities $q=q\left( x^{\gamma },l\right) $ will in general not be
if the gauge change involves $x^{4}=l$. The situation is analogous to that
in quantum field theory, where a choice of gauge (in some cases even a
non-covariant one) is necessary in order to calculate physical
quantities. In the present case, we have two mathematically
acceptable metrics which
have physically different cosmological constants: (17a) has $\Lambda =\left(
3\diagup L^{2}\right) l^{2}\diagup \left( l-l_{0}\right) ^{2}$ and (17b) has
$\Lambda =3\diagup L^{2}$. The latter is standard, insofar as $\Lambda $
is a true constant, which with its astrophysically-indicated size implies $
L\simeq 1\times 10^{28}$ cm \cite{1}. The former is non-standard, because $
\Lambda $ is expected to change as $l$ changes, and can indeed be unbounded
for a certain value of the extra coordinate. On the other hand, if
physics takes place on a hypersurface of constant $l$, then $\Lambda$ is
constant in any case; however, in the induced-matter theory the
observed value of $\Lambda$ depends on the evolution in 5D as determined
by the geodesic equation.

To investigate this in more detail, we will adopt the approach used
elsewhere, in which $l=l\left( s\right) $ is given by a solution of the 5D
geodesic equation \cite{2,8,9}. To shorten the present discussion, we note
that 5D geodesic motion generally implies departures from 4D geodesic motion
(the pure-canonical metric is an exception), and that 5D null paths can
correspond to 4D timelike paths (so a higher-dimensional ``photon'' can
appear as a massive particle in spacetime). To proceed, we return to
the general form of the metric (1), 
for which the
5D geodesic equation splits naturally into a 4D part and an extra part:
\begin{subequations}\begin{align}
\begin{split}&\frac{d^{2}x^{\mu }}{ds^{2}}+\Gamma _{\alpha \beta
}^{\mu }\frac{dx^{\alpha }
}{ds}\frac{dx^{\beta }}{ds} =f^{\mu },   \\
&\quad f^{\mu } \equiv \left( -g^{\mu \alpha }+\frac{1}{2}\frac{dx^{\mu }}{ds}
\frac{dx^{\alpha }}{ds}\right) \frac{dl}{ds}\frac{dx^{\beta }}{ds}\frac{
\partial g_{\alpha \beta }}{\partial l} ,  \end{split}\\
&\frac{d^{2}l}{ds^{2}}-\frac{2}{l}\left( \frac{dl}{ds}\right) ^{2}+\frac{l}{
L^{2}} =-\frac{1}{2}\left[ \frac{l^{2}}{L^{2}}-\left( \frac{dl}{ds}\right)
^{2}\right] \frac{dx^{\alpha }}{ds}\frac{dx^{\beta }}{ds}\frac{\partial
g_{\alpha \beta }}{\partial l} .
\end{align}\end{subequations}
In these, following (17a) and the preceding discussion of metrics, we
substitute
\begin{equation}
g_{\alpha \beta }\left( x^{\gamma },l\right) =\left( \frac{l-l_{0}}{l}
\right) ^{2}k_{\alpha \beta }\left( x^{\gamma }\right) \;\;\;\;,
\end{equation}
where $k_{\alpha \beta}$ is any admissible 4D vacuum metric of general
relativity with a cosmological constant $3 / L^2$.
Furthermore we assume a null 5D path as noted above, and rewrite the line
element as
\begin{equation}
dS^{2}=\left[ \frac{l^{2}}{L^{2}}-\left( \frac{dl}{ds}\right) ^{2}\right]
ds^{2}=0\;\;\;\;.
\end{equation}
Since a massive particle in spacetime has $ds^{2}\neq 0$, we have that the
velocity in the extra dimension is give by $\left( dl\diagup ds\right)
^{2}=\left( l\diagup L\right) ^{2}$. Then the right-hand side of (18b)
disappears, and to obtain the $l$-motion we need to solve
\begin{equation}
\frac{d^{2}l}{ds^{2}}-\frac{2}{l}\left( \frac{dl}{ds}\right) ^{2}+\frac{l}{
L^{2}}=0
\end{equation}
and $(dl/ds)^2=(l/L)^2$ simultaneously. Substituting the latter in
(21), we find that $l$ is a superposition of simple hyperbolic
functions.
There will be two arbitrary constants of integration involved in this
   solution, which can be written as
\begin{equation}
l=A\cosh\left( \frac{s}{L}\right) +B\sinh\left( \frac{s}{L}\right)
\;\;\;\;.
\end{equation}
Moreover, $(dl/ds)^2=(l/L)^2$ implies that $A^2=B^2$. To fix the
constants $A$ and $B$ here, it is necessary to make a choice of
boundary conditions. It seems most natural to us to locate the big bang at
the zero-point of proper time and to choose $l=l_{0}\left( s=0\right)
$. Then $A=l_{0}$ and, $
B=\pm l_{0}$  in (22), which thus reads
\begin{equation}
l =l_{0}e^{\pm s\diagup L }.
\end{equation}
The sign choice here is trivial from the mathematical perspective, and
merely reflects the fact that the motion is
reversible. However, it is not trivial from the physical perspective,
because it changes the behaviour of the cosmological constant.

This is given by (16), which with (23) yields
\begin{equation}
\Lambda =\frac{3}{L^{2}}\frac{1}{(1-e^{\mp s\diagup L})^2}.
\end{equation}
In the first case (upper sign), $\Lambda $ decays from an unbounded
value at the big bang
$\left( s=0\right) $ to its asymptotic value of $3\diagup L^{2}\left(
s\rightarrow \infty \right) $. In the second case (lower sign),
$\Lambda $ decays from
an unbounded value $\left( s=0\right) $ and approaches zero $\left(
s\rightarrow \infty \right) $. We infer from astrophysical data \cite{1} that
the first case is the one that corresponds to our universe.

To investigate the physics further, let us now leave the last component of
the 5D geodesic (18b) and consider its spacetime part (18a). We are
especially interested in evaluating the anomalous force per unit mass $
f^{\mu }$ of that equation, using our metric tensor (19). The latter gives $
\partial g_{\alpha \beta }\diagup \partial l=2\left( l-l_{0}\right) \left(
l_{0}\diagup l^{3}\right) k_{\alpha \beta }\left( x^{\gamma }\right)
=2\,l_{0}\left[ l\left( l-l_{0}\right) \right] ^{-1}g_{\alpha \beta }$ in
terms of itself. We can substitute this into (18a), and note that the
4-velocities are normalized as usual via $g_{\alpha \beta }\left( dx^{\alpha
}\diagup ds\right) \left( dx^{\beta }\diagup ds\right) =1$. The result is
\begin{equation}
f^{\mu }=-\frac{l_{0}}{l\left( l-l_{0}\right) }\frac{dl}{ds}\frac{dx^{\mu }}{
ds}.
\end{equation}
This is a remarkable result. It describes an acceleration in spacetime
which depends on the 4-velocity of the particle and whose magnitude (with
the choice of boundary conditions noted above) is infinite at the big
bang. It is typical of the non-geodesic motion found in other
applications of
induced-matter and membrane theory \cite{5,6}. It follows from (23) that
\begin{equation}f^\mu =\mp
\frac{1}{L}\frac{dx^\mu}{ds}\frac{1}{(e^{\pm s\diagup L}-1)}.
\end{equation}
In the first case (upper sign), $f^{\mu }\rightarrow \left( -1\diagup
s\right) \left(
dx^{\mu }\diagup ds\right) $ for $s\rightarrow 0$ and $f^{\mu }\rightarrow 0$
for $s\rightarrow \infty $. In the second case (lower sign), $f^{\mu
}\rightarrow
\left( -1\diagup s\right) \left( dx^{\mu }\diagup ds\right) $ for $
s\rightarrow 0$ and $f^{\mu }\rightarrow \left( -1\diagup L\right) \left(
dx^{\mu }\diagup ds\right) $ for $s\rightarrow \infty $. Thus both cases
have a divergent, attractive nature near the big bang. However, at late
times the acceleration disappears in the first case, but persists (though is
small if $L$ is large) in the second case. As in our preceding discussion
of $\Lambda $, we infer from astrophysical data on the dynamics of galaxies
\cite{1} that the first case is the one that corresponds to our universe.

As regards anomalous accelerations, let us consider the implications of (26)
above. That equation shows that there is an extra force (per unit mass)
which is proportional to both the velocity in the extra dimension $\left(
dl\diagup ds\right) $ and the velocity in spacetime $\left( dx^{\mu }\diagup
ds\right) $. The first dependency shows that the extra force arises from
motion with respect to the extended coordinate frame, so it is inertial in
the Einstein sense (like centrifugal force). The second dependency shows
that the acceleration is coupled to the dynamics in 4D. In fact,
there is a kind of restoring force
towards the rest state. This is of importance for the dynamics of
galaxies, for it shows that the comoving frame used in standard 4D
cosmological models is actually an
equilibrium state. This agrees with data which show that the Hubble flow
is smooth and that the peculiar velocities of galaxies are small at the
present epoch \cite{1,2}. However, while there are constraints from data on
the 3K microwave background, some dynamical departures must have been
present at early epochs. This with (26) opens a new perspective on the
problem of the formation of structure in the early universe. The standard
theory, wherein a small statistical perturbation of the density is supposed
to grow by gravitational instability into a galaxy, has long been known to
suffer from a timescale problem. The basis of this is that a small
perturbation does not have enough gravitational pull to counteract the rate
at which matter is being diluted by the Hubble expansion, thus limiting the
rate at which it can grow. To explain the properties of galaxies as they
are observed, the process needs to happen faster. The anomalous
acceleration (26) of our model may resolve this problem, since it augments
gravity and thereby assists galaxy formation.

\section{Conclusion and Discussion}

It is apparent from the contents of the preceding section that the
cosmological ``constant'' may not be what it appears to be. In this
section, we review the foregoing algebraic results, and then summarize the
physical consequences of what we have found.

The metric (1) of general relativity extended to five dimensions can in
principle handle all of 5D physics. However, it is instructive to look at
the restricted case of (4D) matter-free conformally-flat metrics (2), as they are the
analogs of the inflationary de Sitter cosmology. The field equations for
apparently empty 5D space are given by (3), and these are known by Campbell's
theorem to contain all solutions of the 4D Einstein equations \cite{2}. The
latter involve the cosmological ``constant'', which can either be regarded
as related to an extra force in addition to gravity, or as a measure of the energy
density of the vacuum. The
pure-canonical 5D metric $(\partial g_{\alpha\beta}/\partial l=0)$ yields $\Lambda =3\diagup L^{2}$, where $L$ is a
length that scales the 4D part of the metric, and which is known from
astrophysical data to be $L\simeq 1\times 10^{28}$ cm \cite{1}. A simple gauge
transformation, wherein the extra coordinate is shifted by a constant,
revalues $\Lambda $ to (16), which is variable if there
is motion in the extra dimension. This is constrained by the 5D geodesic
equation, which splits naturally into a 4D part involving an extra force
(18a) and an extra part in $x^{4}=l$ (18b). The motion in the extra
dimension can easily be found (23), which enables $\Lambda =\Lambda \left(
s\right) $ to be evaluated (24). The motion in the regular dimensions of
spacetime can likewise be evaluated, but involves an extra force (per unit
mass) or acceleration (26).

The main physical result of our working is that in 5D, the cosmological
constant is changed from $\Lambda =3\diagup L^{2}$ to $\Lambda =\left(
3\diagup L^{2}\right) l^{2}\left( l-l_{0}\right) ^{-2}$, where $l$ is the
value of the extra coordinate. This is the result of merely changing $l$
to $\left( l-l_{0}\right) $. Such a result may appear at first sight to be
surprising, but in retrospect it could have been foreseen: If we extend
general relativity from 4 to 5 dimensions, {\it any} change in the
extra coordinate will preserve the 5D formalism but alter the 4D one.
Covariance is powerful, and if applied in $N\left( \geq 5\right)$ D
will
alter our view of 4D physics. In other words, if the world has more than 4
physically-significant dimensions, what we perceive in spacetime depends on
how we choose the gauge (coordinate frame) in 5 dimensions. The results we
have found can be viewed as a test of whether or not there are more than 4
dimensions: The decay of the cosmological ``constant'' as in (24) and the
existence of a fifth force as in (26) are both in principle open to
test. These are small effects as measured at the present epoch, and
in conformity
with current astrophysical data \cite{1}. But both of these effects must,
according to the present model, have drastically influenced the early
universe. Galaxy formation, in particular, must have been influenced by an
anomalous acceleration that complements the decay of $\Lambda $ and augments
gravity.

If we can obtain such significant physical effects from a mere shift along
the axis of a minimally-extended version of general relativity, it is
permissible to wonder about more complicated gauge changes.
Phenomenological arguments have recently been made which indicate
that a
particle may have its ``own'' associated $\Lambda $, connected to its mass
(see, e.g., \cite{Man}). Our results are the consequence of a particularly simple change of
gauge.

\section*{Acknowledgements}

This work grew out of an earlier collaboration with H. Liu. It was
supported by University of Missouri-Columbia and N.S.E.R.C. (Canada).

\end{document}